\documentstyle[epsfig,psfig,rotate]{aa}

\begin{document}
\thesaurus{08(02.19.1;              
	      09.01.1;              
	      09.09.1 Orion Peak 1; 
	      09.13.2;              
	      12.03.3;              
	      13.09.4)              
}
\title{
Detection of HD in the Orion molecular outflow \thanks{
Based on observations with ISO, an ESA project with instruments funded
by ESA Member States (especially the PI countries: France, Germany, The
Netherlands and the United Kingdom) and with the participation of ISAS
and NASA.
}
}
\author{
Frank Bertoldi\inst{1}\inst{2}    \and
Ralf Timmermann\inst{2}   \and
Dirk Rosenthal\inst{2}    \and
Siegfried Drapatz\inst{2} \and Christopher M. Wright\inst{3}\inst{4}
}
\institute{
Max-Planck-Institut f\"{u}r Radioastronomie,
Auf dem H\"ugel 69, D-53121 Bonn, Germany, bertoldi@mpifr-bonn.mpg.de
\and
Max-Planck-Institut f\"{u}r extraterrestrische Physik,
Giessenbachstrasse, D-85740 Garching, Germany
\and
Leiden Observatory, P.O.~Box~9513, NL-2300 RA~Leiden, The Netherlands
\and
School of Physics, University College, UNSW, 
ADFA, Canberra ACT 2600, Australia,
wright@ph.adfa.edu.au
}

\offprints{F. Bertoldi}
\date{Received 5 November 1998; accepted 18 February 1999}
\authorrunning{F. Bertoldi et al.}
\maketitle

\begin{abstract}

We report a detection in the interstellar me\-di\-um of an infrared
transition within the electronic ground state of the deuterated
hydrogen molecule, HD.  Through a deep integration with the
Short-Wavelength-Spectrometer (SWS) on board the Infrared Space
Observatory (ISO), the pure rotational $v=0-0$ R(5) line at
19.43~$\mu$m was detected toward the Orion (OMC-1) outflow at its
brightest H$_2$\ emission region, Peak 1.  The $\sim20''$ beam-averaged
observed flux of the line is $(1.84 \pm 0.4) \times 10^{-5}\rm ~erg
~cm^{-2}$ $\rm s^{-1} sr^{-1}$.  Upper flux limits were derived for
sixteen other rotational and ro-vibrational HD lines in the wavelength
range 2.5 to 38~$\mu$m.

We utilize the rich spectrum of H$_2$\ lines observed at the same
position to correct for extinction, and to derive a total warm HD
column density under the assumption that similar excitation conditions
apply to H$_2$\ and HD. Because the observed HD level population is not
thermalized at the densities prevailing in the emitting region, the
total HD column density is sensitive to the assumed gas density,
temperature, and dissociation fraction.
Accounting for non-LTE HD level populations in a partially dissociated
gas, our best
estimate for the total warm HD column density is $N({\rm HD})=(2.0\pm
0.75)\times 10^{16}~{\rm cm^{-2}}$.
The warm molecular hydrogen column density is
$(2.21\pm0.24)\times 10^{21}~{\rm cm^{-2}}$, so that the
relative abundance is
$\rm [HD]/[H_2]=(9.0\pm 3.5)\times 10^{-6}$.

The observed emission presumably arises in the warm layers of
partially dissociative magnetic shocks, where HD can be depleted relative to
H$_2$ due to an asymmetry in the deuterium-hydrogen exchange reaction.
This leads to an average HD depletion relative to H$_2$\ of about
40\%. Correcting for this chemical depletion, we derive a deuterium
abundance in the warm shocked gas, [D]/[H]= $(7.6\pm 2.9)$ $\times
10^{-6}$.

The derived deuterium abundance is not very sensitive to the dissociation
fraction in the emitting region, since both the non-LTE and the chemical
depletion corrections act in oppositite direction.
Our implied deuterium abundance is low compared to previous determinations
in the local interstellar medium, but it is consistent with two other recent
observations toward Orion, suggesting that deuterium may be significantly
depleted there.


\end{abstract}

\section{Introduction}
\label{se:introduction}

Deuterium is an important clue to the physics of the Big Bang. Its
creation rate during primordial nucleosynthesis
depended strongly on the number ratio of photons to baryons, a quantity not at
all well-known but crucial for a correct description of the earliest events
(e.g., Wilson \& Rood 1994; Smith et al. 1993).
Ever since,
conditions have not been right to add further deuterium to the
primordial production; neither nuclear fusion processes nor the spallation of
heavier nuclei by energetic cosmic rays can augment the original abundance
(although stellar flares were suggested by Mullan \& Linsky
[1999] to produce deuterium).
The deuterium abundance in fact decreases continuously as deuterium is
burned up in stars.
The present day deuterium abundance in the interstellar medium
provides a lower limit to its
primordial value, and it
reflects the history of stellar reprocessing of the gas;
measurements of its spatial variation may shed light on
the star formation history in a given region (e.g., Tosi 1998,
Tosi et al. 1998).

Deuterium and D-bearing species were detected in giant planets
(Encrenaz et al. 1996; Feuchtgruber et al. 1997), in cosmic rays
(Beatty et al. 1985; Casuso \& Beckman 1997; Geiss \& Gloeckner 1998),
in the local
ISM (Linsky et al. 1993, 1995; Linsky 1998; Piskunov et
al. 1997; Helmich et al. 1996; Jacq et
al. 1993; Turner 1990), and in extragalactic sources (Songaila et
al. 1994; Tytler et al. 1996; Burles \& Tytler 1998).  Through recent
\ion{D}{i} Ly$\alpha$ observations of the local ISM, McCullough
(1992), Linsky et al. (1993, 1995), Piskunov et al. (1997), and
Dring et al. (1997) derived
a deuterium abundance [D]/[H]~$\simeq 1.5 \times 10^{-5}$ along the
line of sight toward nearby
bright stars. These observations are hampered however
by the possible confusion with the \ion{H}{i} Ly$\alpha$ line, which
is only $\simeq$80~km~s$^{-1}$\ apart.
High spectral resolution UV absorption
observations of DI Lyman $\delta$ and $\epsilon$
were recently performed by Jenkins et al. (1999) and Sonneborn et al.
(in prep.) along three
lines of sight, which yield significant [D]/[H] variations between
$7.4\times 10^{-6}$ and $2.1\times 10^{-5}$.
Observations of the hyperfine transitions of \ion{D}{i}
and \ion{H}{i} at radio frequencies (Chengalur et al. 1997), or of the
H$_2$\ and HD molecules, provide further means to sample the
deuterium abundance.
The [HD]/[H$_2$] ratio is 
subject to variations due to small differences in chemical
reaction rates in high temperature molecular gas
as in shocks (see Sect.~\ref{se:chemistry}) and photodissociation regions.
To date, rotationally excited HD has been
detected with ISO
from giant planets (Feuchtgruber et al. 1997),
and in the interstellar medium
through UV-observations by Wright \& Morton (1979).
Accounting for the three lowest rotational levels of HD
Wright \& Morton found
[HD]/[H$_2$]~$< 6 \times 10^{-7}$ in the cold molecular gas
toward $\zeta$ Oph. D-bearing species such as HDO, CH$_2$DOH,
DCO$^+$ or DCN (Jacq et al. in prep.), are
observed at radio-frequencies in the ISM, but due to fractionation the
deuterium abundance cannot be deduced accurately from these observations.

The infrared emissivity of HD has been modeled and predicted for
photodissociation regions (Sternberg 1990 - not including the
reaction $\rm H+HD\getsto H_2+D$) and magnetic (C-type)
shocks (Timmermann 1996), where latter calculations include deuterated
species in the chemical reaction network.  But even
state-of-the-art IR-spec\-tro\-meters had been unsuccessful in detecting
HD emission in the ISM, because of the low HD abundance, and because the pure
rotational and rotation-vibrational lines appear at wavelengths that are
affected by strong telluric interference.

ISO now for the first time opened the skies to a successful search for
HD emission.  Wright et al.~(1999) detected the emission of HD 0-0
R(0) toward the Orion Bar photodissociation region, and derive
[HD]/[H$_2$]$= (2.0\pm 0.6)\times 10^{-5}$. A measure of the
deuterium abundance in PDRs is complicated by the fact
that the HD dissociation front is located deeper into the molecular
cloud than that of H$_2$.
The average excitation of HD may therefore be lower than that
of H$_2$, making it difficult to derive column densities referring to
the same regions. In regions of pure collisional excitation such as
shocks, this problem should not occur.

We here report HD observations toward the brightest H$_2$\ emission
region in the sky, Peak~1 in the Orion
molecular outflow that surrounds several deeply embedded 
far-infrared sources
(Genzel \& Stutzki 1989; Menten \& Reid 1995; Blake 1997;
van Dishoeck et al. 1998; Stolovy et al. 1998; Schultz et al. 1998).
Our observations, which we discuss in Sect.~2, 
were part of a line survey of shocked
molecular gas, and of a program to investigate the oxygen-chemistry in
the warm molecular interstellar medium with the
Short-Wavelength-Spectrometer (SWS).
In Sect.~3 we derive an HD column density from
the flux of one detected HD line, making use also of extinction 
and excitation measurements from a large number of  H$_2$\
lines which we detected in related observations.
We then estimate the deuterium abundance in the
shocked, warm gas of the Orion outflow. 
In Sect.~4 we summarize our results.


\section{Observations}
\label{se:observations}

%
\begin{figure*}
\begin{minipage}[t]{0.71\linewidth}
\centering
\includegraphics[width=11cm]{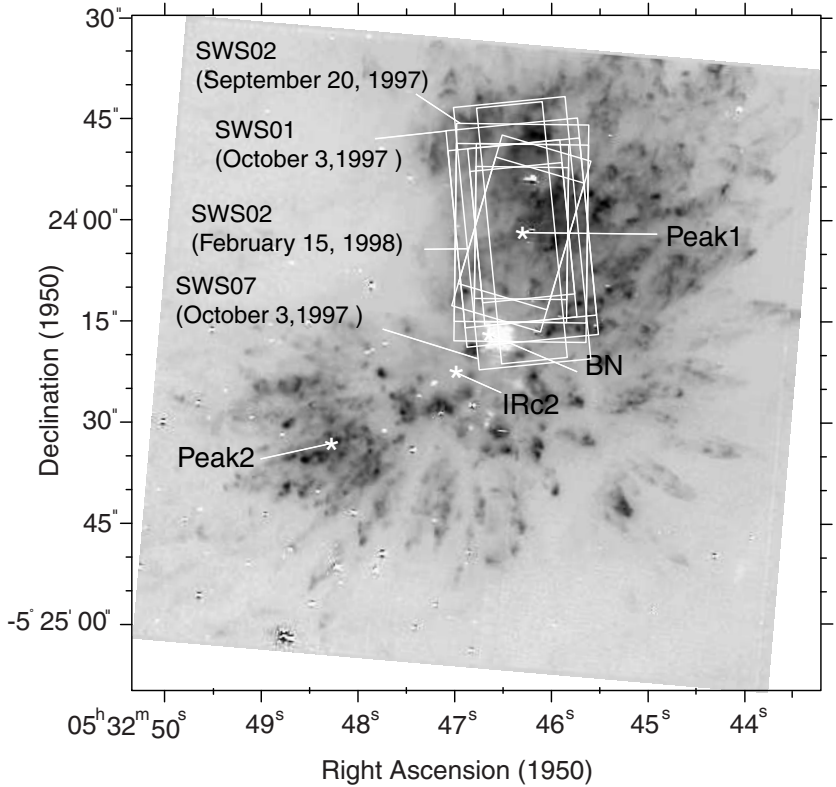}
\end{minipage}
\begin{minipage}[t]{0.24\linewidth}
\centering
\vspace*{-10.5cm}
\caption{
The various SWS apertures of our ISO observations of the Orion outflow,
overlaid
on a NICMOS H$_2$\ 1--0 S(1) map kindly provided by E.~Erickson and A.~Schultz
(Schultz et al. 1998).
White dots and the central white patch are continuum ghost-images of
stars and of Orion-BN, respectively.
The ISO aperture was centered on
$\rm 05^h 32^m 46^s.27$, $-05\degr ~24\arcmin~ 02\arcsec$ (1950).
Its
size corresponds to $14\arcsec \times 20\arcsec$ for wavelengths
smaller 12 $\mu$m, $14\arcsec \times 27\arcsec$ at 12 to 27.5 $\mu$m,
$20\arcsec \times 27\arcsec$ at 27.5 to 29 $\mu$m, and $20\arcsec
\times 33\arcsec$ at 29 to 45.2 $\mu$m.}
\label{fig:peak1}
\end{minipage}
\end{figure*}

\begin{table*} 
\caption[]{
Summary of ISO-SWS observations of HD lines at Orion Peak~1. 
}
\begin{tabular}{crcrcrrrrr}
\hline\noalign{\smallskip}

\multicolumn{1}{c}{transition} & 
\multicolumn{1}{c}{$\lambda^{~~a}$} & 
\multicolumn{1}{c}{integration$^b$} & 
\multicolumn{1}{c}{peak} &
\multicolumn{1}{c}{$\Delta v~^c$} &
\multicolumn{1}{c}{$A$} &
\multicolumn{1}{c}{$E_{vJ}$} &
\multicolumn{1}{c}{$I_{\rm obs}$} &
\multicolumn{1}{c}{$A_{\lambda}$} &
\multicolumn{1}{c}{$N_{vJ}$$^d$} \\ 
\multicolumn{1}{c}{ } & 
\multicolumn{1}{c}{$\mu$m}    & 
\multicolumn{1}{c}{sec}        &      
\multicolumn{1}{c}{Jy} & 
\multicolumn{1}{c}{km~s$^{-1}$} & 
\multicolumn{1}{c}{s$^{-1}$} &
\multicolumn{1}{c}{K} &
\multicolumn{1}{c}{erg~s$^{-1}$cm$^{-2}$sr$^{-1}$} &
\multicolumn{1}{c}{mag} &
\multicolumn{1}{c}{cm$^{-2}$} \\ 

\noalign{\smallskip}
\hline\noalign{\smallskip}
0--0 R(2) & 37.7015 &      & $<$ 192  & 300 & 1.72(--6)$^e\!\!\!$ & 765.9& $<$9.9(--4)& 0.09 & $<$ 1.5(17)\\
0--0 R(3) & 28.5020 &      & $<$ 100  & 350 & 4.11(--6) & 1270.7 & $<$ 7.9(--4)        & 0.30 & $<$ 4.6(16)\\
0--0 R(4) & 23.0338 &      & $<$ 35   & 350 & 7.91(--6) & 1895.3 & $<$ 6.0(--4)        & 0.51 & $<$ 1.8(16)\\
0--0 R(5) & 19.4305 & 3010 & 2.37     & 134 & 1.33(--5) & 2635.8 & $(1.84\pm0.4)$(--5) & 0.60 & $(3.0\pm 1.1)$(14)\\
0--0 R(6) & 16.8940 &      & $<$ 8.5  & 230 & 2.03(--5) & 3487.5 & $<$ 1.3(--4)        & 0.53 & $<$ 1.1(15)\\
0--0 R(7) & 15.2510 & 1110 & $<$ 5.0    & 185 & 2.88(--5) & 4445.3 & $<$ 6.8(--5)        & 0.41 & $<$ 3.3(14)\\
0--0 R(8) & 13.5927 &      & $<$ 10   & 215 & 3.87(--5) & 5503.8 & $<$ 1.8(--4)        & 0.37 & $<$ 5.6(14)\\
0--0 R(9) & 12.4718 &      & $<$ 8   & 230 & 4.97(--5) & 6657.5 & $<$ 1.7(--4)        & 0.54 & $<$ 4.3(14)\\
1--0 P(4) & 3.0690  &      & $<$ 0.2 & 280 & 7.37(--6) & 5958.5 & $<$ 2.8(--5)        & 1.11 & $<$ 2.0(14)\\
1--0 P(3) & 2.9800  &      & $<$ 0.1  & 270 & 1.09(--5) & 5593.5 & $<$ 1.4(--5)        & 1.08 & $<$ 6.4(13)\\
1--0 P(2) & 2.8982  &      & $<$ 0.25 & 300 & 1.64(--5) & 5348.7 & $<$ 3.9(--5)        & 0.94 & $<$ 1.0(14)\\
1--0 P(1) & 2.8225  &      & $<$ 0.4  & 305 & 3.23(--5) & 5225.7 & $<$ 6.6(--5)        & 0.80 & $<$ 7.6(13)\\
1--0 R(0) & 2.6900  &      & $<$ 0.2  & 330 & 1.72(--5) & 5348.7 & $<$ 3.7(--5)        & 0.70 & $<$ 7.0(13)\\
1--0 R(1) & 2.6326  &      & $<$ 0.3  & 330 & 2.51(--5) & 5593.5 & $<$ 5.7(--5)        & 0.70 & $<$ 7.3(13)\\
1--0 R(2) & 2.5811  &      & $<$ 0.4   & 245 & 3.22(--5) & 5958.5 & $<$ 5.8(--5)        & 0.72 & $<$ 5.7(13)\\
1--0 R(3) & 2.5350  & 1200 & $<$ 0.2 & 215 & 3.91(--5) & 6441.1 & $<$ 2.6(--5)        & 0.74 & $<$ 2.1(13)\\
1--0 R(4) & 2.4943  &      & $<$ 0.35 & 235 & 4.60(--5) & 7037.7 & $<$ 5.0(--5)        & 0.76 & $<$ 3.5(13)\\

\noalign{\smallskip}        \hline
\noalign{\smallskip}
\end{tabular}

$^{a}$ expected wavelengths. 0-0 R(5) is observed at $19.4290~\mu$m. \\
$^{b}$ on target integration time for SWS 02 observations; the SWS~01
2.4 -- 40 $\mu$m scan took 6538 sec. \\
$^{c}$ FWHM of observed HD line, or of neighboring H$_2$ lines.\\
$^{d}$ upper level column, corrected for extinction.\\
$^{e}$ numbers in brackets denote powers of ten.\\
\label{tab:lines}
\end{table*} 

Orion Peak~1 was observed in the SWS~01 (full grating scan) and SWS~07
(Fabry-P\'erot) modes of the short wavelength spectrometer (de Graauw
et al. 1996) on board ISO (Kessler et al. 1996) on October 3, 1997,
and in the SWS~02 ($\approx 0.01~\lambda$ range grating scan)
mode on September 20, 1997 and February 15, 1998.
Figure \ref{fig:peak1}
illustrates the various aperture orientations with respect to the H$_2$\
1--0 S(1) emission observed with NICMOS on the HST (Schultz et
al. 1998).  The full 2.3 to 45~$\mu$m SWS~01 spectrum was recorded in
its slowest mode with the highest possible resolution.  A preliminary
reduction of this spectrum was presented by Bertoldi (1997).
Table~\ref{tab:lines} summarizes the HD line observations. The H$_2$\
lines will be discussed in more detail in a forthcoming article
(Rosenthal et al. in prep.).

The data reduction was carried out using standard Off Line Processing
(OLP) routines up to the Standard Processed Data (SPD) stage within
the SWS Interactive Analysis (IA) system. Between the SPD and Auto
Analysis Result (AAR) stages, a combination of standard OLP and
in-house routines were used to extract the individual spectra. The
in-house routines included an interactive dark-current subtraction for
individual scans and detectors as well as for the removal of
fringes. The flux calibration errors range from 5\% at 2.4~$\mu$m\ to
30\% at 45~$\mu$m.  (SWS-Instrument Data User Manual, version 3.1). The
statistical uncertainties derived from the line signal to noise ratio
are for most detected lines smaller than the systematic errors due to
flux calibration uncertainties.

\subsection{Line Fluxes}
\label{se:flux}

\begin{figure*} 
\resizebox{\hsize}{!}{\includegraphics{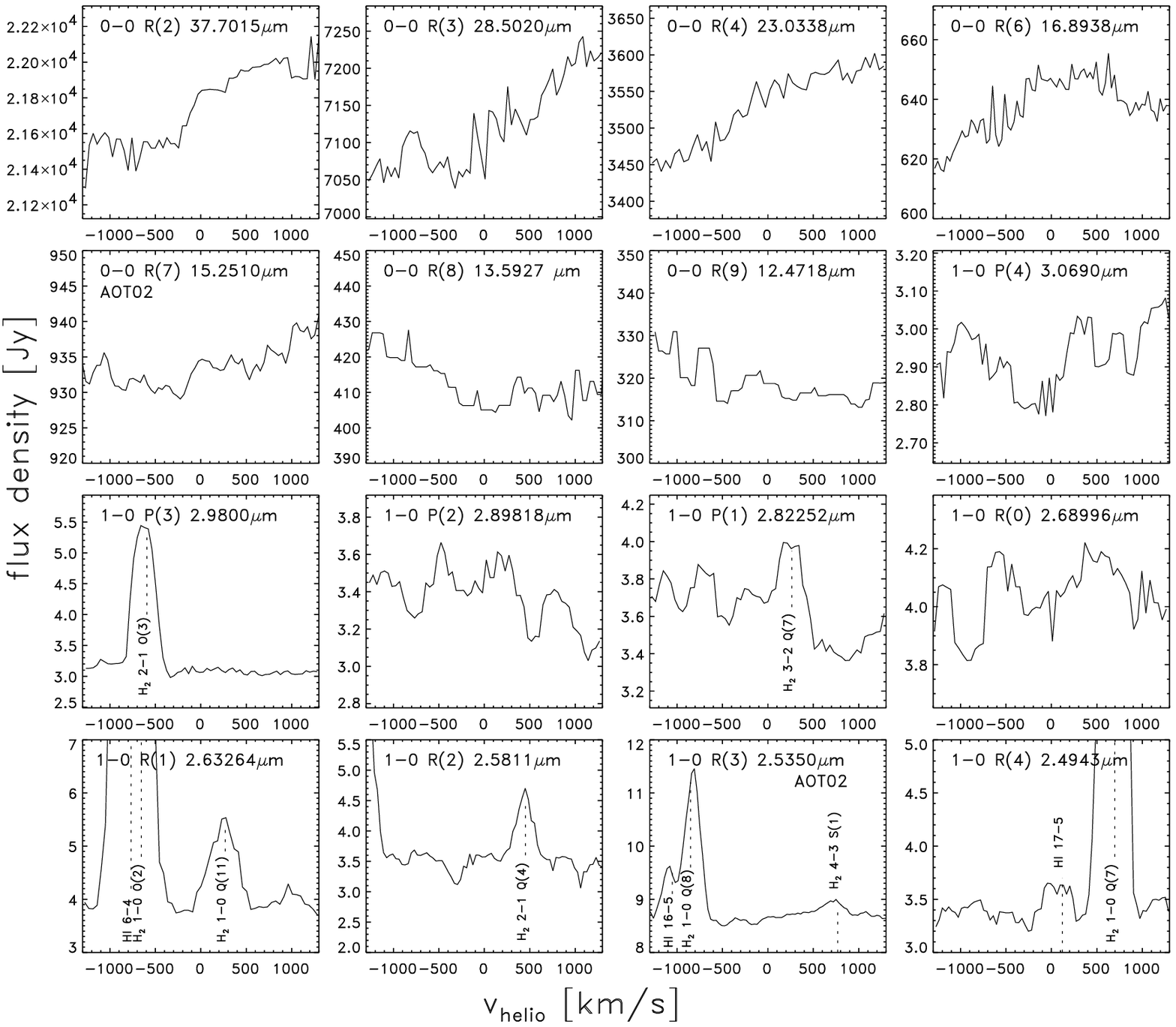}}
\vspace{-0.9cm}
\caption{
Spectra of HD non-detections. 
Three of the spectra were obtained in the
SWS~02 mode with ISO, all others were taken in the SWS~01 mode.
Neighboring H$_2$\ and \ion{H}{I}
lines are identified at their expected positions.
}
\label{fig:hd_spec}
\end{figure*} 

\begin{figure} 
\epsfig{file=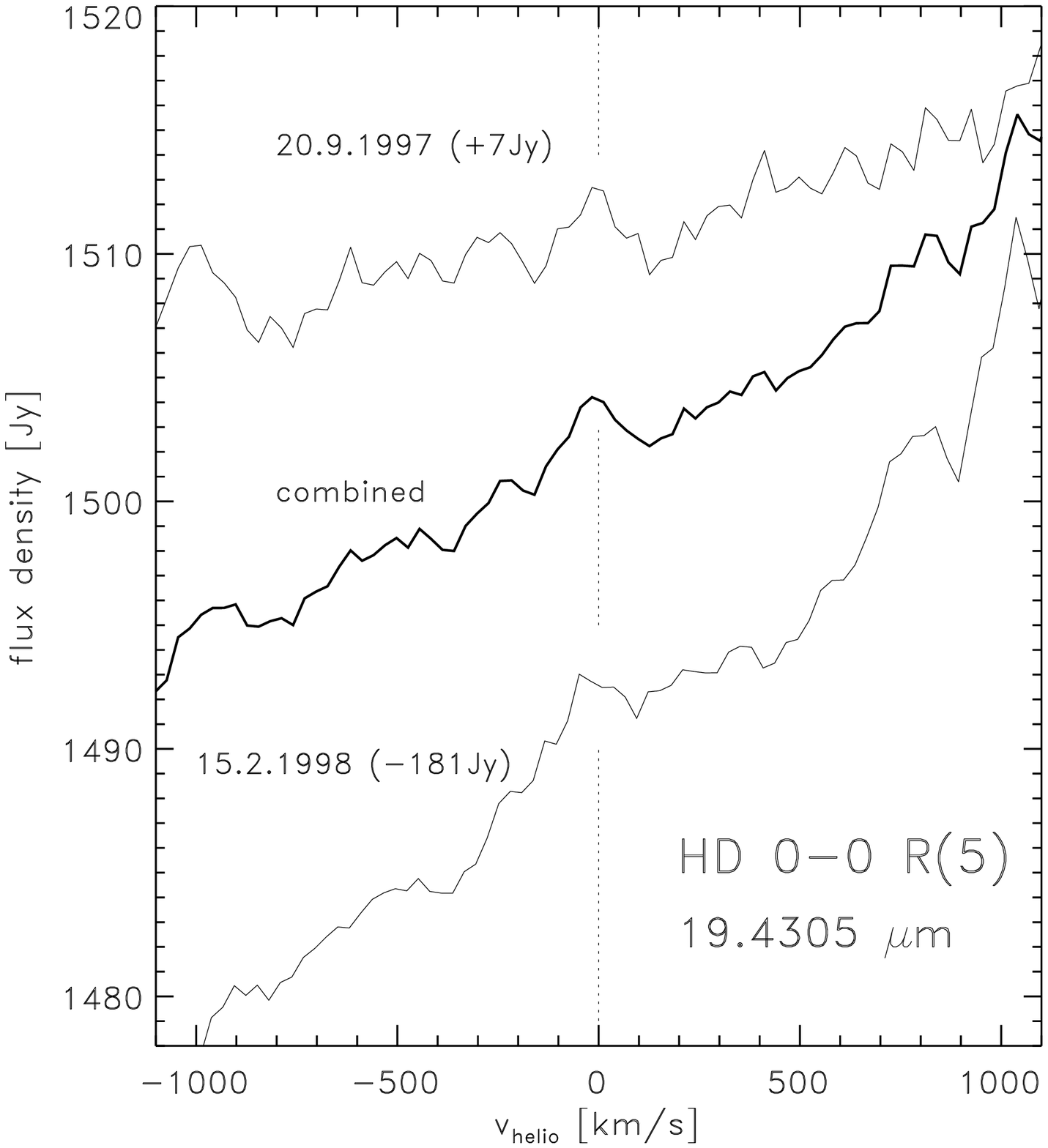,width=8.7cm,
clip=,bbllx=2pt,bblly=12pt,bburx=497pt,bbury=555pt}
\caption{Defringed AOT~02 spectra of two 0--0 R(5) observations performed at the
labeled dates, and the merged spectrum.
The line is visible even in the
original fringed data of both observations. Continuum levels
differed in both datasets (due to differing orientations of the aperture
with respect to Orion-BN), so that a constant was added to one before averaging
them, and in order to display them together.
The rising noise near the scan edges is expected and due to the lower
detector coverage. In the merged spectrum, the line stands out with
a peak flux density to RMS noise ratio of 4.5.
$v_{\rm helio}$ refers to the expected line wavelength of 19.4305 $\mu$m.}
\label{fig:00R5}
\end{figure}

The spectra at the expected wavelengths of seventeen HD transition lines
are shown in Figs.~\ref{fig:hd_spec} and \ref{fig:00R5}. 
The wavelengths
(Table \ref{tab:lines}) of the $v=0-0$ pure rotational HD transitions
were adopted from Ulivi et al. (1991), while those for the
rotation-vibrational lines were computed from the rotational
(Essenwanger \& Gush 1984) and vibrational 
constants (Herzberg 1950).  Only one of the seventeen lines
sought was detected, the 0--0 R(5) (i.e. $J=6\rightarrow 5$) dipole
transition at an expected wavelength of 19.4305~$\mu$m\footnote{
The experimental value for the R(5) transition wavelength is
$19.4305\pm0.0001\mu$m, whilst the calculated value is 
$19.431002\pm0.000008\mu$m (Ulivi et al. 1991 and references therein).
The discrepancy is yet unexplained.
}
(observed at
19.4290 $\mu$m, see Fig.~\ref{fig:00R5}).  
Although the detection may seem marginal, the line does
appear in two independent observations. It is apparent in the
fringed data, and after defringing,
it stands out well above the noise in the
central wavelength range of the coadded SWS~02 scans.
A beam-averaged flux of $(1.84 \pm 0.4) \times
10^{-5}$erg~s$^{-1}$cm$^{-2}$sr$^{-1}$\ was derived
from a continuum-subtracted integration over the feature;
the error derives from 
the line's S/N$=4.5$ (the RMS noise being evaluated within
$\pm 500~{\rm km~s^{-1}}$)
plus an estimated 11\% flux calibration uncertainty.
The line  width (FWHM)
is 134~km~s$^{-1}$, which agrees with the instrumentally expected width
for extended objects, 130~km~s$^{-1}$, and is not sufficient to resolve
the emission, which should have a velocity dispersion similar
to that observed in H$_2$ 1--0 S(1), which has a
FWHM of $\approx 50$ km~s$^{-1}$, with emission from -100 to +100 km s$^{-1}$
(Chrysostomou et al. 1997; Stolovy et al. 1998).
The line center is positioned at
$v_{\rm helio} \simeq -23$~km~s$^{-1}$, which is within the
range of line center
velocities, $v_{\rm helio} \simeq -38$ to +41~km~s$^{-1}$, observed in the
H$_2$\ 1--0 S(1) emission (Chrysostomou et al. 1997). The high
spectral resolution and sensitivity of the ISO SWS 
around 19~$\mu$m made $0-0$ R(5) the
most promising line for a detection of HD in Peak 1. The stronger lines
from lower
rotational states suffer from the rapidly rising
continuum level at longer wavelengths and the resulting strong fringing.

The available 
integration time was insufficient to detect two other HD lines
which we tried to observe in the SWS~02 mode.
We derived upper flux limits for a total of
sixteen HD lines from the noise level near the expected line
wavelengths (Table~\ref{tab:lines}).  The peak-to-peak noise envelope
approximately corresponds to a 3$\sigma$ flux density dispersion, and
we assumed that a line with a peak flux density of 3$\sigma$ would
have stood out clearly enough to be detected. As upper flux limits we
therefore adopted the 3$\sigma$ flux density noise level
times the FWHM of neighboring H$_2$\ lines, except for 0--0 R(7), R(8),
R(9), for
which we adopted the linewidths expected from instrumental resolution.

\section{Extinction correction and H$_2$\ excitation}
\label{se:excitation}

\begin{figure} 
\epsfig{file=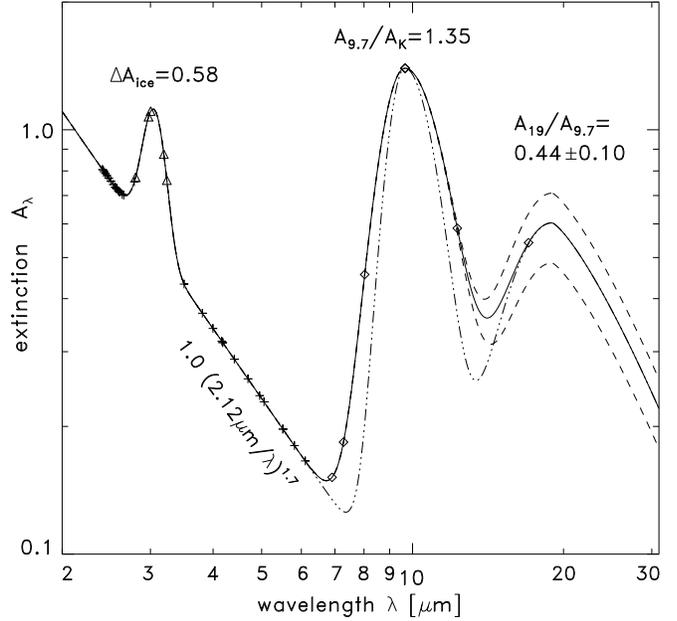,width=8.7cm,
clip=,bbllx=8pt,bblly=28pt,bburx=490pt,bbury=485pt}
\caption{
Adopted extinction curve (solid line), including water ice and silicate
features.
Dashed lines show curves with different relative silicate features strengths,
$A_{19}/A_{10}= 0.35$ and 0.52. The dash-dotted curve assumes narrower
9.7~$\mu$m
feature. Wavelenghts of those
H$_2$ lines used to constrain the extinction curve are
marked -- but these are not datapoints!}
\label{fig:extinction}
\end{figure}

The column density of molecules in a particular rotation-vibrational level
$(v,J)$ is computed from the observed  line flux,
$I_{obs}(v,J\rightarrow v',J')$,  of a transition to
a lower state $(v',J')$, times
the wavelength, $\lambda$, divided by the radiative transition rate
(Einstein) coefficient\footnote{
	The Einstein coefficients for H$_2$\ and HD are taken
	from Turner et al.~(1977), confirmed by Wolniewicz et
	al. (1998), and Abgrall et al. (1982), respectively. The
	transition energies for H$_2$\ were computed from the level
	energies that were kindly provided by Roueff (1992, private
	communication).}
for this transition, $A(v,J\rightarrow v',J')$, times a correction for
extinction along the line of sight:
	\begin{equation}
	N_{vJ} ~=~ { 4 \pi\over h c}~{ \lambda~ I_{obs}(v,J\rightarrow v',J')
	\over A(v,J\rightarrow v',J') }
	~~10^{0.4 A_{\lambda}}~,
	\end{equation}
where $A_{\lambda}$ is the effective extinction at wavelength $\lambda$.  
 All H$_2$\ and HD transitions are optically thin.
Previous measurements of the Peak 1 molecular emission
estimated K band (2.12~$\mu$m) extinctions between 0.5 and 1 magnitude
(Everette et al. 1995).
To apply an extinction correction to the observed H$_2$\ and HD lines 
requires knowledge of the ``extinction law" $A_\lambda/A_{\rm K}$ between
2.4 and 40~$\mu$m, which especially above $\sim 5$~$\mu$m
is  observationally not well constrained
(Draine 1989). The near-IR extinction is usually approximated
to follow a power law
(see Fig.~\ref{fig:extinction}),
	\begin{equation}
	A_\lambda = A_{\rm K} \; 
	\left(~\lambda~/~2.12~ \mu{\rm m}~\right)^{-\epsilon}
	~~~~~~~{\rm for}~~\lambda < 6~\mu{\rm m},
	\label{eq:dust}
	\end{equation}
with an increase in opacity beyond 6~$\mu$m due to stretch and bend mode
resonances in silicate grains (Draine \& Lee 1984), the shape and
depth of which are yet poorly understood.  A recent
ISO study of the extinction toward the W51 \ion{H}{ii} region, e.g.,
finds $A_{19}/A_{10}\simeq 0.52 \pm 0.1$ and $A_{19}/A_{\rm K}\simeq
0.57\pm 0.1$
(Bertoldi et al. in prep.). Draine \& Lee (1984) suggested
$A_{18}/A_{9.7}\simeq 0.40$, and observations of circumstellar dust
emissivities suggest values for this ratio 
between 0.35 (Pegourie \& Papoular 1985) and 0.5 (Volk \& Kwok 1988).

The effective extinction toward the emitting gas in Peak 1 may not
necessarily follow an average interstellar extinction law, since the
emitting and absorbing gas may be mixed.  We therefore tried to
estimate the effective extinction as a function of wavelength toward
Peak 1 from the observed emission of Peak 1 itself. The differential
extinction between two wavelengths can be derived from a comparison
of H$_2$\ line fluxes of transitions arising either from the same upper
level, or from neighboring thermalized levels with a well determined
relative excitation.  Since the warm HD and H$_2$\ are likely to be well
mixed, we will use the extinction toward the H$_2$\ to deredden also the
HD line intensities.

The extinction-corrected H$_2$\ column density distribution serves as a
``thermometer'' that probes the excitation conditions in the emitting
gas.  Since we detected only one HD line, we cannot determine the HD
excitation directly --this would require at least two lines of
transitions arising from different upper levels. Instead, we rely on
the reasonable assumption that HD is subject to the same excitation
conditions as H$_2$. We can thereby derive both molecules' total column
densities in the warm, emitting gas, and from that, their abundance
ratio.
 
\subsection{2.4 -- 6~$\mu$m extinction}
\label{se:dust}

It is useful to plot the observed H$_2$\ or HD column densities, $N_{vJ}$,
divided by the state's statistical weight,\footnote{
	The statistical weights are $g_{J}=
	2J+1$ for para-H$_2$\ (even $J$) and $g_{J} = 3(2J+1)$ for 
	ortho-H$_2$ (odd $J$).
	HD has non-identical nuclei (and a small dipole moment
	$\sim 10^{-4}$~Debye). Therefore,
	unlike H$_2$, which exists in the form
	para-H$_2$\ and ortho-H$_2$, such a difference does not exist for
	HD, and	$g_J=2J+1$.}
$g_J$, against the state's energy, $E_{vJ}$.  Examining this
``excitation diagram,'' which we first 
plotted from the not yet extinction-corrected H$_2$\
line intensities, we found no indication up to $E_{vJ}/k\approx
40,000$~K of
fluorescent excitation, or for deviations from the statistical ortho-
to para-H$_2$\ abundance ratio of 3 (details will be given by
Rosenthal et al. in prep.).  Instead, the
populations appear both in thermodynamic and statistical
equilibrium, which means, they depend only on the level energy
and degeneracy, not on the vibrational quantum number. As a consequence,
$N_{vJ}/g_J$ is a smooth function of the level energy.

In contrast, H$_2$\ that is fluorescently excited  
generally displays level populations that are not
in vibrational {\it and} rotational LTE,
and also show ortho-to-para column density
ratios between neighboring states that are smaller than 3 (see e.g., Draine
\& Bertoldi 1996, Bertoldi 1997).  Timmermann (1998) predicted
that in low-velocity shocks the ortho-to-para ratio can also be 
significantly lower, which was confirmed in recent ISO-SWS
observations by Neufeld et al. (1998), who found a value of
$\simeq 1.2$ in the outflow HH54. Peak 1 however shows no deviations
from statistical equilibrium.

\begin{figure} 
\psfig{file=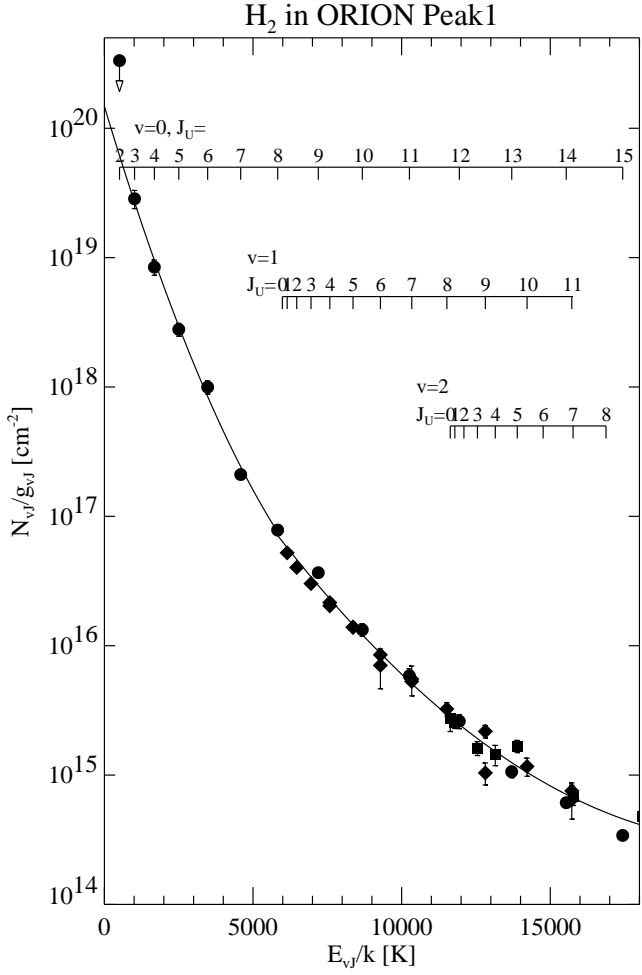,width=8.7cm, 
clip=,bbllx=45pt,bblly=70pt,bburx=520pt,bbury=790pt}
\caption{
H$_2$\ excitation diagram: the dots, diamonds, and squares denote
dereddened and beam-averaged column densities of transitions
in the $v$ = 0--0, 1--0, and 2--1 bands, respectively. The solid line
represent the least squares fit, Eqs.~(\ref{eq:fit2}) and (\ref{eq:fit1}), 
which apply for $E_{vJ}/k$ larger and smaller than 5800~K, respectively.}
\label{fig:h2_ex}
\end{figure}

We were able to obtain the most reliable line fluxes 
for the H$_2$\ $v=0-0$ and
$v=1-0$ lines between 2.4 and 6~$\mu$m that have upper level energies
$E_{vJ}/k = 5000 - 16,000$~K. Excluding lines in the water
ice absorption feature between 2.8 and 3.3~$\mu$m, 
we selected thirty-two high S/N lines
in this wavelength and energy range to constrain the power law part of
the extinction curve.
Assuming that the H$_2$\ columns
$N_{vJ}/g_J$ vary smoothly with $E_{vJ}$, we fit a
second order polynomial to the observed column densities,  
corrected for extinction
following Eq.~(1) with $A_{\rm K}$ and $\epsilon$ as free parameters.  We
searched for values of $A_{\rm K}$ and $\epsilon$ which minimize the
dispersion of $N_{vJ}/g_J$.  
Within the range $\epsilon= 1-2$,
$A_{\rm K}$ is well constrained to $1.0\pm 0.1$ magnitudes. It was not
possible to narrow down the value of $\epsilon$ further, so we adopted
$\epsilon=1.7$, which was suggested by previous observational
studies (see Draine 1989; Brand et al. 1988 used $A_{\rm K}=0.8$
and $\epsilon=1.5$).
The second order fit to $N_{vJ}/g_J$ versus $E_{vJ}/k\equiv 1000~ T_3$~K
for $A_{\rm K}=1.0$ and $\epsilon=1.7$ is (see Fig.~\ref{fig:h2_ex})
	\begin{equation}
	\log(N_{vJ}/g_J)~ =~ 18.88 - 0.402 \: T_3 + 0.0092 \: T_3^{2},
	\label{eq:fit2}
	\end{equation}
for $5800{\rm ~K}<E_{vJ}/k<17,000$~K, and has a fit quality $\chi^2=0.13$;
the uncorrected ($A_{\rm K}=0$) dispersion is $\chi^2=0.39$.

We should note that our extinction curve between 4 and 7 $\mu$m\ is not
constrained well enough to address the claim by Lutz et al. (1997)
that the extinction curve flattens and lacks the presumed minimum at
7 $\mu$m.

\subsection{Mid-IR extinction}
\label{se:mid-IR}

Having constrained the extinction curve below 6~$\mu$m and the H$_2$\
excitation from 5800 -- 17,000~K, we now estimate the strength of the
10~$\mu$m silicate feature with the observed H$_2$\ $v=0$, $J\le 8$ column
densities.  
The intensity of the 0--0 S(3) line at 9.66~$\mu$m gives a
column density of the $v=0$, $J=5$ level that is significantly 
below that expected from an interpolation between the $J=3$ 
(0--0 S(1) 17.0~$\mu$m) and $J=8$ (0--0 S(6) 6.1~$\mu$m) levels.
The observed $J=3$ to $J=7$ level populations are all affected by silicate
absorption. We estimate the shape of the 9.7~$\mu$m and 18~$\mu$m
silicate features from the calculations of Draine \&
Lee (1984), and adjust the strengths of the features
individually (Fig.~\ref{fig:extinction}). 
Assuming $A_{\rm K}=1.0$ mag and tuning $A_{9.7}$ 
such that the $J=5$ level column density smoothly
follows that of the $J=3-8$ levels (Fig.~\ref{fig:h2_ex} shows the
resulting dereddened columns) results in a possible range  $A_{9.7} =
(1.3-1.5)$~mag, if we allow for a variation in
$A_{19}/A_{10}=0.35-0.52$, and a 15\% uncertainty of 
the $J=5$ column due to the difference between two 
measurements of the
0--0 S(3) line in different AOTs. When we narrow the 10~$\mu$m
feature below reasonable width estimates, the extinction correction for
the $J=4$ and $J=6$ levels decreases significantly, 
thereby lowering the dereddened flux
in the 0--0 S(3) line and decreasing the required extinction
to $A_{9.7}\approx (1.1-1.3)$~mag.

Considering all uncertainties we estimate $A_{9.7}\approx
(1.35\pm 0.15)$~mag, and with $A_{19}/A_{10}\approx 0.35 - 0.52$, we
find $A_{19} \approx (0.61 \pm 0.15)$~mag. Our value for
$A_{19}/A_{\rm K} = (0.61 \pm 0.15)$ is in good agreement with the
$0.57 \pm 0.1$ found toward W51. 

The extinction correction for the detected HD
line we then estimate as $10^{0.4 A_{19}} \approx 1.75 \pm 0.25$. 
The dereddened column densities of the H$_2$\ $v=0$, $J\le 8$ levels are
fit by 
	\begin{equation}
	\log\left(N_{vJ}/g_J\right) ~=~ 20.17 - 0.765 \: T_3 +
	0.0344 \: T_3^{2}.
	\label{eq:fit1}
	\end{equation}
At upper level energies
$E_{Jv} / k = 0$, 2636~K (HD $J=6$) and 5800~K this corresponds to
excitation temperatures $T_{ex} \simeq 570$, 740 and 1190~K, respectively. 

\subsection{Water ice feature}
\label{se:water}

The rotation-vibrational HD transitions 1--0 P(2), P(3), and P(4)
fall into the water ice absorption feature at $3.05\pm 0.25$~$\mu$m. 
To correct these line flux limits for extinction 
we estimated the depth and width of the feature from the apparently
enhanced (over the power law, Eq.~[\ref{eq:fit2}]) extinction of
five H$_2$\ lines
with $6000{\rm ~K}<E_u/k<\rm 16,000$~K and 2.8~$\mu$m~$<\lambda<$~3.3~$\mu$m. 
The Gaussian  
	\begin{equation}
	\Delta A_{ice}(\lambda)~ \approx ~0.58
	~e^{-[(\lambda-3.05{\rm \mu m})/(\sqrt2\cdot 0.15{\rm \mu m})]^2}~ 
	{\rm mag}
	\end{equation}
approximately fits the additional 
extinction noticeable for the five H$_2$\ lines, and was used
to correct the rovibrational HD line fluxes.

\begin{figure}
\psfig{file=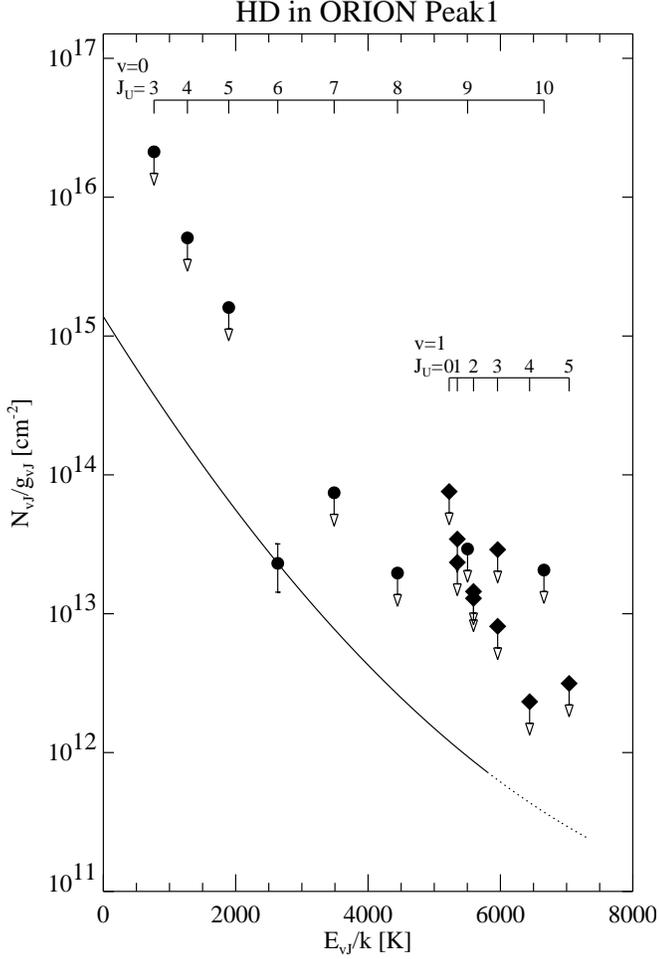,width=8.7cm, 
clip=,bbllx=45pt,bblly=60pt,bburx=525pt,bbury=780pt}
\caption{
HD excitation diagram. Pure rotational transitions are denoted by
dots, and $v$ = 1--0 transitions by diamonds. The
line represents the fit Eq.~(\ref{eq:fit3}). 
The error of the 0--0 R(5) line
is computed from spectral noise (22\%) and
uncertainties in the flux calibration (11\%)
and the extinction at 19.4~$\mu$m ($\simeq$14\%).
}
\label{fig:hd_ex}
\end{figure}

\subsection{Total column of $\rm H_2$ and {\rm HD}}
\label{se:LTE}

How can we derive the total HD column density with only one measured
high-excitation level?  We can make use of the observed H$_2$\
excitation, and estimate that the excitation conditions -- i.e., the
fractions of the total gas column density that are at a particular
temperature and density -- are the same for HD and for H$_2$. We assume
that at least up to the excitation energy of HD $J=6$
($E_{06}/k=2636$~K), the H$_2$\ level populations are thermalized, and
thereby reflect the kinetic temperature distribution of the gas.  
The assumption of thermalized level populations is
supported by the lack of
deviations from vibrational degeneracy or non-statistical
ortho-to-para ratios, and by detailed non-LTE
calculations of the H$_2$\ level populations, which show that deviations
from LTE are small at $J\le 8$ ($E/k\le 5800$~K) 
for densities higher than $10^5~{\rm cm^{-3}}$ and
temperatures above 600~K (Draine \& Bertoldi, unpublished).

Because they are permitted dipole
transitions, the radiative decay rates
of HD are much higher than those of H$_2$\
levels at comparable energy. 
This results in  ``critical'' densities (i.e. densities 
above which a level is
thermalized for a given temperature) which are higher for HD than for H$_2$.
If the gas density was high enough to thermalize the
HD levels, then HD would show the same level excitation as H$_2$:
normalizing
the H$_2$\ populations (\ref{eq:fit1}) with the measured HD $J=6$ level,
we would then expect the HD populations to follow
	\begin{equation}
	\log(N_{vJ}/g_J) ~=~ 15.14 - 0.765 ~ T_3 + 0.0344 ~ T_3^2~
	\label{eq:fit3}
	\end{equation}
for $T<6000$~K. 
The measured upper limits to sixteen HD level column densities
are consistent with this distribution (Fig.~\ref{fig:hd_ex}).

However, deviations from LTE are important at the
expected density of $10^5-10^6~{\rm cm^{-3}}$ and temperatures of $600-1000$~K in the
shocked gas of the OMC-1 outflow.
Let $n_{vJ}/n_{{\rm LTE},vJ}$ be the actual non-LTE 
population in level $(v,J)$, divided
by its LTE value.
The total warm
HD column density is given by the sum over all level column densities
	\begin{equation}
	N({\rm HD}) ~=~ \frac{n_{\rm LTE,06}}{n_{06}} ~
	\sum_{vJ} \left(N_{vJ}\over g_J\right)
	~(2J+1) ~ \frac{n_{vJ}}{n_{{\rm LTE},vJ}},
	\label{eq:HDtot}
	\end{equation}
where $N_{vJ}/g_J$ is given by Eq.~(\ref{eq:fit3}).

\begin{figure} 
\epsfig{file=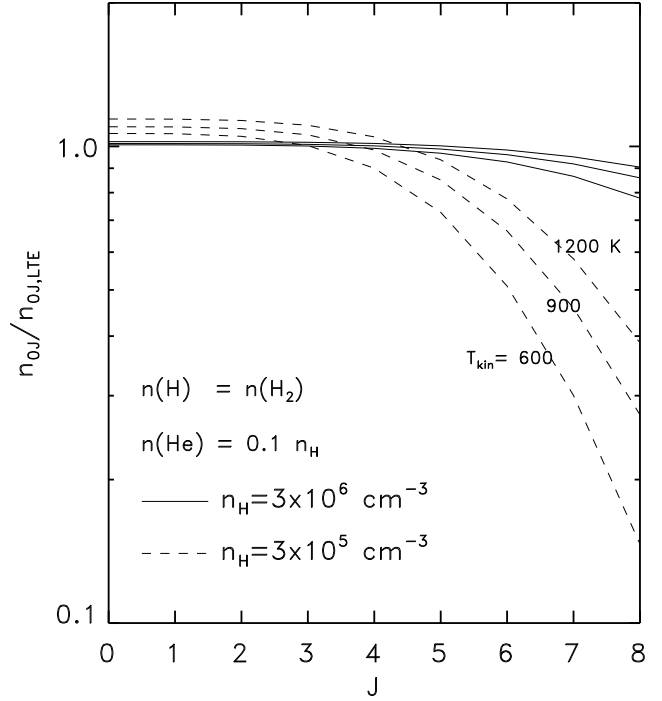,width=8.7cm,
clip=,bbllx=35pt,bblly=25pt,bburx=440pt,bbury=465pt}
\caption{
Deviations from LTE HD level populations plotted against the rotation
quantum number $J$ for different densities and kinetic gas
temperatures. $n_{0J}$ and $n_{{\rm LTE},0J}$ refer to densities in an
individual $v=0$ state under non-LTE and LTE
conditions.
}
\label{fig:hd_pop}
\end{figure} 

To compute the deviations from LTE of the HD level population we solved the
equations of statistical equilibrium (see Timmermann 1996), including
radiative decay and excitation/deexcitation through collisions with He,
H$_2$, and H. The abundance of H varies throughout the shock, and we
adopted [H]/[H$_2$]$=1$, which is typical for the dissociation fraction
in the hot layers of a partially dissociative shock.
In Fig.~\ref{fig:hd_pop} we compare the resulting level
population to that in LTE for gas volume densities of $3\times 10^5$ and
$3\times 10^6{\rm cm^{-3}}$, and three different kinetic temperatures that should
span the range typical of the observed, shocked gas. 

HD collisional (de)excitation rate coefficients with H$_2$ were
computed properly only for pure rotational transitions up to $J=4$ and
temperatures up to 600 K and 300 K, respectively (Sch\"afer 1990).
The rate coefficients for higher level transitions and temperatures were
extrapolated from the lower ones. The H--HD and H$_2$--HD collision rate
coefficients were recently computed by Roueff \& Flower (1999) and
Roueff \& Zeippen (1999), respectively.

We find that for H$_2$\ densities below $10^6~{\rm cm^{-3}}$ the
HD level populations 
deviate significantly from LTE. At a density of $3\times 10^5~{\rm cm^{-3}}$
and a temperature 
of 600~K, the population of the $J=6$ level is about a factor two
below its LTE value.

\begin{figure} 
\epsfig{file=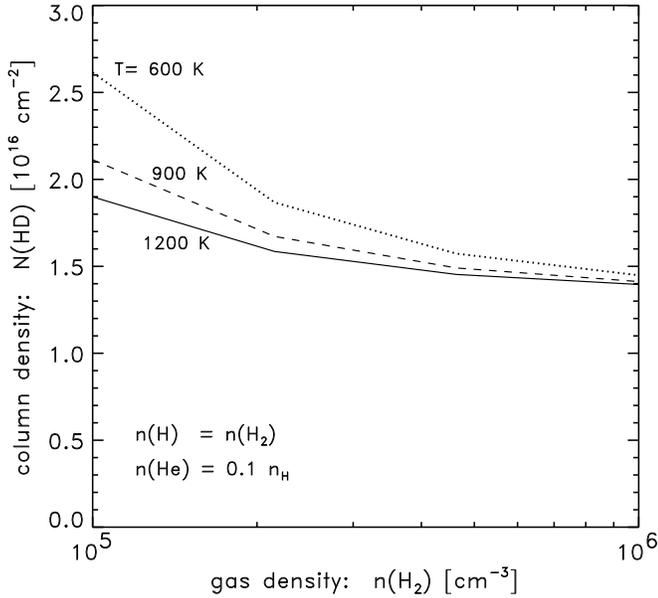,width=8.7cm,
clip=,bbllx=10pt,bblly=20pt,bburx=495pt,bbury=465pt}
\caption{Total HD column density computed from non-LTE level populations
as a function of H$_2$\ density for three different gas temperatures.
The H$_2$\ density in the preshock gas has been 
estimated at $(1-3.5)\times 10^{5}\rm cm^{-3}$, and the
shocked gas temperatures at $\approx 600-1000$~K, so we estimate $N({\rm HD})
\approx (2.0\pm0.75)\times 10^{16}\rm cm^{-2}$.
}
\label{fig:NHD}
\end{figure} 

To assess the non-LTE effects on the derived total HD column density,
we evaluated $N({\rm HD})$ from Eq.~(\ref{eq:HDtot})
and plot this in Fig.~\ref{fig:NHD} as a function
of the gas density for three different temperatures.
Only at H$_2$\ densities above $10^6~{\rm cm^{-3}}$ 
do the populations assume LTE, and the warm HD column is
	\begin{equation}
	N({\rm HD})_{\rm LTE} ~=~ (1.36\pm 0.38)\times 10^{16}\rm cm^{-2}~.
	\end{equation}
The error derives from the uncertainty of the $J=6$ column
measured by the 0--0 R(5) line, and is a combination of
spectral noise (22\%), flux calibration uncertainty (11\%),
and the uncertainty of the extinction correction (14\%). 
At lower densities, the total column density
much depends on the gas temperature,
and since H is the strongest collision partner of HD,
on the dissociation fraction.

The emission of the OMC-1 outflow
had been modeled previously with C-shocks that propagate at velocities
of order $35-40~{\rm km~s^{-1}}$ into gas with
densities of $n({\rm H_2})=(1-3.5)\times 10^5~{\rm cm^{-3}}$
(Draine \& Roberge 1982; 
Chernoff et al.~1982; Kaufman \& Neufeld 1996). 
The gas in
such shocks reaches temperatures of order 1000~K,
and remains at approximately constant density through the region where
most of the H$_2$ emission occurs.
The lowest H$_2$\ levels we observed 
show excitation temperatures of 600--700~K, which probably
reflects the kinetic temperature of much of the warm, emitting gas.
Taking these temperatures and densities as estimates for the prevailing 
excitation conditions, we estimate that the total observed 
HD column density must be in the range
	\begin{equation}
	N({\rm HD}) ~=~ (2.0\pm 0.75)\times 10^{16}\rm ~ cm^{-2}~.
	\end{equation}
If we had neglected H--HD collisions, the HD level would be less thermalized,
and our estimate for the total HD column would rise to
$(3.5\pm 1.4)\times 10^{16}\rm ~ cm^{-2}$. However, we believe that
much of the emission arises from partially dissociative shocks, so that
collisions with neutral hydrogen should be important.

We now compare the HD column to that of warm H$_2$.
Summing over the H$_2$\ level populations given by 
the least squares fits, Eqs.~(\ref{eq:fit2}) and 
(\ref{eq:fit1}), we find
	\begin{equation}
	N({\rm H_2})~=~(2.21\pm 0.24)\times 10^{21}\rm~ cm^{-2}~,
	\end{equation}
where the error reflects a maximum flux calibration uncertainty of 11\%
in the 7 to 19.5~$\mu$m range. By summing from $J=0$, we
extrapolated the observed H$_2$\ level populations, $J\ge 3$, to the
unobserved levels $J=0-2$. We thereby account only for the
{\it warm} H$_2$, not for
the {\it total}~ H$_2$\ column along the line of sight,
which includes over
$10^{22}\rm ~cm^{-2}$ of cold gas in the molecular cloud that
embeds the outflow.
In this cold gas, which we are not concerned with, most H$_2$\ is 
in its ground states $J=0$ and $J=1$, and does 
not affect our analysis of the 
warm outflow gas seen in the higher levels.

Dividing the HD and H$_2$\ column densities, we derive a first estimate of the
abundance ratio
	\begin{equation}
	{\rm [HD]/[H_2]}~=~(9.0\pm 3.5)\times 10^{-6}~.
	\end{equation}

\subsection{Chemical depletion of HD in shocks}
\label{se:chemistry}

\begin{figure} 
\psfig{file=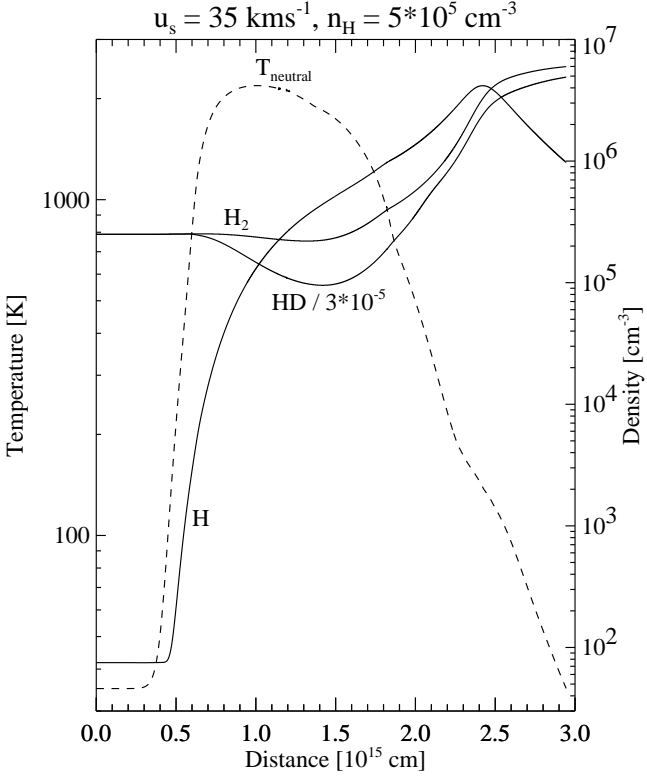,width=8.7cm, height=11.cm,
clip=,bbllx=45pt,bblly=90pt,bburx=535pt,bbury=710pt}
\vspace{-0.5cm}
\caption{Example of the density and temperature structure
in a planar magnetic C-type shock, which was
modeled with the code of Timmermann (1996, 1998). The
pre-shock densities $n({\rm H_2}) = 2.5 \times 10^5\rm ~cm^{-3}$ and $n({\rm
HD}) = 7.5$~cm$^{-3}$, the shock speed $u_{\rm s} =
35$~km~s$^{-1}$, and $B_0=700\mu$G.
The depletion of HD relative to H$_2$\ reaches a maximum
value of 2.4 behind the hottest layer of the shock.
}
\label{fig:hd_shock}
\end{figure} 

The OMC-1 outflow emission arises from warm
molecular gas that is shock-heated by 
high-velocity ejecta which originates from one of 
the deeply embedded protostars in the vicinity of Orion IRc2. 
The emission is 
a mixture from fast, dissociative J-type shocks, in which the molecular emission
comes from where molecules reform, and from slower
($<50\rm ~km~s^{-1}$), partially- or non-dissociative C-shocks, in which
the molecules radiate in a magnetic precursor where
the temperature rises to its peak value (Fig. \ref{fig:hd_shock}).
In J-shocks that propagate into a medium with
density of order $10^5-10^6\rm cm^{-3}$, the fraction of the bulk motion
energy radiated away in H$_2$ lines is very small, of
order 0.1\% (Neufeld \& Dalgarno 1989). 
In C-type shocks, however, about half the kinetic energy is converted
to H$_2$ emission. For J-type shocks to dominate the H$_2$ and HD
emission, about 1000 times more energy would have to be dissipated 
in fast J-shocks than in the slower C-shocks. For an even
distribution of magnetic field strengths and
shock velocities up to $\sim 100~$km~s$^{-1}$, C-shocks therefore
vastly dominate the molecular emission.
The outflow emission may therefore be well modeled as arising from such
magnetic, partially-dissociative shocks.

Most of the deuterium is locked in HD in the dense pre-shock 
molecular gas. In a high-velocity C-shock, both HD and H$_2$\ are  
either dissociated through collisions with ions
that stream through the neutral gas at a speed of order the shock velocity,
or through collisions with warm H$_2$\ and H at temperatures in excess of
2500~K. In a partly dissociative C-shock, 
HD is depleted more than H$_2$\ because atomic hydrogen 
can efficiently destroy HD through the reaction
	\begin{equation}
	 {\rm D + H_2}\getsto {\rm HD + H} + \Delta H_0,
	\label{eq:hddest}
	\end{equation}
where $\Delta H_0/k = 418$~K is the enthalpy
difference. Figure \ref{fig:hd_shock}
illustrates the density profiles for H$_2$\ and HD across such a C-shock.
In the warmest layers, HD is depleted by up to a factor 2.4
relative to H$_2$. In the cooling region, the atomic
deuterium reacts stronger again with H$_2$\ to form HD,
and the equilibrium of Eq.~(\ref{eq:hddest}) is shifted back
towards HD. Averaged over the region where the
temperature exceeds 400~K, from where
most of the observable emission arises,
HD is depleted relative to H$_2$\ by a factor 1.67 in the particular case
we display in Fig.~\ref{fig:hd_shock}.


In C-shocks propagating at velocities below 25~km~s$^{-1}$, H$_2$\ is not
dissociated significantly, and because of the low abundance 
of atomic hydrogen, the chemical depletion of HD is negligible.

\subsection{Deuterium abundance}
\label{se:abundance}

To derive the deuterium abundance in the warm, shocked gas 
we adopt the non-LTE level distribution of HD we computed above, 
and further account for
the possibility of chemical depletion. With the column density ratio
derived for temperatures of 600--900~K and densities of
$(1-3.5)\times 10^5~{\rm cm^{-3}}$, we found 
	\begin{equation}
	{\rm [D]/[H]= 0.5 [HD]/[H_2]} = (4.5 \pm 1.7)\times 10^{-6}~, 
	\end{equation}
and accounting for chemical depletion of HD by 1.67, the abundance
is raised to 
	\begin{equation}
	{\rm [D]/[H]}= (7.6 \pm 2.9)\times 10^{-6}~, 
	\end{equation}
where no error was added for uncertainties in the chemical
depletion factor.

The main uncertainty of our value appears to arise from 
the indirect measure of the HD excitation, and of the abundance of atomic
hydrogen in the warm shocked region. Less dissociative shocks
would require larger corrections for non-LTE level populations, which
would raise the implied HD abundance: neglecting all H--HD collisions,
we earlier derived $N({\rm HD})\approx (3.5\pm 1.4)\times 10^{16}\rm cm^{-2}$,
which yields [D]/[H]$=7.9\times 10^{-6}$. Now chemical depletion should not
occur in shocks with a low abundance of neutral hydrogen, so that this
values reflects the actual deuterium abundance. Interestingly, we find that
the effects of non-LTE and of
chemical depletion nearly cancel, so that the derived deuterium
abundance turns out to be not very sensitive on how dissociative the
shock is.


\section{Summary}

ISO for the first time enabled the detection in the interstellar
medium of an infrared transition in the electronic ground state of
deuterated hydrogen, HD.  We here report the
discovery of the $v$=0--0 R(5) line at 19.4290 $\mu$m
with the ISO Short Wavelength Spectrometer in the warm,
shocked molecular gas of the Orion OMC-1 outflow, at the bright emission 
``Peak 1.''  Upper flux limits for sixteen other HD lines were
measured, all of which appear consistent with expectations when
considering the observed 0--0 R(5) line flux.

A large number of H$_2$\ lines were detected (Rosenthal et al. in prep.) and
utilized to analyze the HD observations. The near- and mid-infrared
extinction toward the emitting region was derived by minimizing the
dispersion in the observed H$_2$\ level column densities with respect to
an LTE excitation model. Thereby we derive a near-infrared
(K band) extinction of $(1.0\pm0.1)$ magnitudes, a 9.7 $\mu$m
extinction of $(1.35\pm 0.15)$ mag, and from an estimated range of
$A_{19}/A_{9.7}= 0.35 - 0.52$ we correct the HD 0-0 R(5) flux for
extinction by $(0.61\pm0.15)$ mag, i.e. a factor $1.75\pm0.25$.

The dereddened H$_2$\ level populations served to estimate the
excitation conditions in the gas. While H$_2$\ was assumed to have
thermalized level populations, those for HD were
computed in detail by making use of the H$_2$\ excitation.
Due to non-LTE effects at $J>3$,
the total warm HD column density was found to
be sensitive to the gas density and temperature at the 
densities estimated to
prevail in the shocked gas,
$n({\rm H_2})\approx(1-3.5)\times  10^5~{\rm cm^{-3}}$. 

Our estimate for the observed warm HD column density is 
$N({\rm HD})=(2.0\pm0.75 )\times 10^{16}~{\rm cm^{-2}}$, and for the warm
molecular hydrogen, $N({\rm H_2})=(2.21\pm0.24)\times 10^{21}~{\rm cm^{-2}}$.
Their relative abundance is therefore $\rm [HD]/[H_2]=(9.0\pm 3.5)\times
10^{-6}$. 

We note that in high-velocity C-shocks, HD may be 
depleted relative to H$_2$ because of an asymmetry (due to a small
binding energy difference) in the
deuterium-hydrogen exchange reaction ${\rm HD + H} \getsto {\rm D + H_2}$.
Estimating that this lowers the warm HD column
by about 40\%, we derive a deuterium abundance
in the warm shocked gas, [D]/[H]$=(7.6\pm2.9)\times 10^{-6}$.

If the emitting shocks were on average less dissociative than assumed, the chemical
depletion would be less pronounced.
But at the same time, the lower H-HD collision rate
would enhance the HD $J=6$ level population's deviation from LTE, to the
effect that our implied
total column of HD would increase. We estimate that the two effects
approximately cancel, and that [D]/[H] remains at $\approx 8\times 10^{-6}$,
independent of how dissociative the shock actually is.

The major uncertainties in our estimate for the deuterium abundance 
arise from our indirect measure of
the HD excitation, and since we must therefore rely on non-LTE excitation
models, on the
uncertainty of the preshock density and the
abundance of neutral hydrogen in the
shock.  Future ground-based near-IR observations of
ro-vibrational transition lines of HD could 
constrain the effects of non-LTE and thereby narrow the error margins
of the deuterium abundance. Detailed shock models of the recently resolved,
multiple 
bow-shaped emission in the Orion outflow (Schultz et al. 1998)
might also yield better
estimates for the pre-shock densities and shock velocities, and thereby
of the H$_2$ dissociation and of the
chemical depletion fraction of HD in such shocks.

The deuterium abundance we find, $(7.6\pm 2.9)\times 10^{-6}$,
is lower than that derived through most DI absorption measurements in
the local ISM, but it is consistent with
that found recently
by Jenkins et al. (1999) toward $\lambda$ Orionis, $(7.4^{+1.9}_{-1.3})
\times 10^{-6}$,
and by Wright et al. (1999) toward the Orion Bar, $(1.0\pm 0.3) \times 10^{-5}$.
This could indicate that toward Orion, the deuterium abundance is indeed
somewhat lower than on average.


We are thankful to J.~Lacy, B.~Draine, and M.~Walmsley
for valuable comments, to A.~Schultz for providing the NICMOS
image, and to the SWS Data Center at MPE, especially to H.~Feuchtgruber
and E.~Wieprecht,
for their support. SWS and ISODC at MPE are supported by DARA under grants
50QI86108 and 50QI94023. FB acknowledges support by the 
Deutsche Forschungsgemeinschaft (DFG) through its ``Physics of
Star Formation'' program. 
CMW acknowledges support by NFRA/NWO grant 781-76-015 and of an ARC Research 
Fellowship.


{ }
\end{document}